\documentclass[useAMS,usenatbib]{mn2e}
\usepackage{graphicx,amssymb,epsfig}

\newcommand{\1}{^{-1}}
\def\Msun{\hbox{$\rm\, M_{\odot}$}}

\newcommand{\changed}[1]{{\bf #1}}

\title[Globular Cluster Formation Within The Aquarius Simulation]{Globular Cluster Formation Within The Aquarius Simulation}
\author[B. F. Griffen et al.]{B. F. Griffen$^{1}$\thanks{E-mail: griffen@physics.uq.edu.au} 
, M. J. Drinkwater$^{1}$
, P. A. Thomas$^{2}$
, J. C. Helly$^{3}$
, K. A. Pimbblet$^{1}$
\\
$^{1}$Department of Physics, University of Queensland, QLD 4072, Australia\\
$^{2}$Astronomy Centre, University of Sussex, Falmer, Brighton BN1 9QH, UK\\
$^{3}$Institute for Computational Cosmology, Dept. of Physics, University of Durham, South Road, Durham DH1 3LE, UK}

\begin{document}

\date{25 September 2009} 

\pagerange{\pageref{firstpage}--\pageref{lastpage}} \pubyear{2009}

\maketitle

\label{firstpage}

\begin{abstract}

The Aquarius project is the first simulation that can resolve the full mass range of potential globular cluster formation sites.  With a particle mass $m_\mathrm{p}=1.4 \times 10^4$\Msun, Aquarius yields more than 100 million particles within the virial radius of the central halo which has a mass of $1.8 \times 10^{12}$\Msun, similar to that of the Milky Way. With this particle mass, dark matter concentrations (haloes) as small as 10$^6$ M$_\odot$ will contain a minimum of 100 particles.

Here, we use this simulation to test a model of metal-poor globular cluster formation based on collapse physics. In our model, globular clusters form when the virial temperatures of haloes first exceed $10^4$\,K as this is when electronic transitions allow the gas to cool efficiently.  We calculate the ionising flux from the stars in these first clusters and stop the formation of new clusters when all the baryonic gas of the galaxy is ionised. This is achieved by adopting reasonable values for the star formation efficiencies and escape fraction of ionising photons which result in similar numbers and masses of clusters to those found in the Milky Way. The model is successful in that it predicts ages (peak age $\sim$ 13.3 Gyrs) and a spatial distribution of metal-poor globular clusters which are consistent with the observed populations in the Milky Way. The model also predicts that less than 5$\%$ of globular clusters within a radius of 100 kpc have a surviving dark matter halo, but the more distant clusters are all found in dark matter concentrations.

We then test a scenario of metal-rich cluster formation by examining mergers that trigger star formation within central gas disks. This results in younger ($\sim$ 7--13.3 Gyrs), more centrally-located clusters (40 metal-rich GCs within 18 kpc from the centre of the host) which are consistent with the Galactic metal-rich population. We test an alternate model in which metal-rich globular clusters form in dwarf
galaxies that become stripped as they merge with the main halo. This process is inconsistent with observed metal-rich globulars in the Milky Way because it predicts spatial distributions that are far too extended.
\end{abstract}


\section{Introduction}


Globular clusters (GCs) provide a remarkably rich source of information about
galaxy formation. Unlike the diffuse stellar populations of galaxies, globular
clusters mostly contain stellar populations with a narrow range of ages
and are extremely homogenous, making them relatively simple
to understand and model. The are extremely old, so they survive as a
record of conditions and processes of the earliest stages of galaxy formation
(\citealt{West04}). The properties of the Milky Way halo GC population
led \cite{Searle78} to conclude that the halo of our Galaxy formed via the
slow accretion of many small proto-galactic fragments, not via monolithic
collapse as previously thought (\citealt{Eggen62}). In the 1990s, the (old) ages of globular
clusters provided one of the motivations for considering cosmological models
with non-zero cosmological constants (\citealt{Ostriker95}).

In the last two decades observations of extragalactic GC populations
(notably with the {\em Hubble Space Telescope}) have revealed strong
bimodality in the optical colours of GCs \citep[e.g.][]{Ashman92}. The
blue population is identified as metal-poor clusters and very old,
whereas the red population is more metal-rich and not as
old. 

These observations have motivated a
number of competing galaxy and GC formation scenarios
\citep[e.g.][]{Forbes97,West04}, which attempt to explain
the bimodal colours, but no conclusive theory has emerged. However, the very existence of the bimodal colours is indicative of two epochs of star formation. In view of
the large amount of data collected, there is a strong need for more
detailed theoretical work that will provide specific predictions of
where and when GCs formed \citep{Ashman98, Brodie06}.

In contrast with the extensive gains from observational studies of GCs, it has
proven very difficult to predict the full range of observed GC properties in a
self-consistent manner from theoretical models. The mass and spatial scales
needed to study the physical conditions of GC formation are very difficult to
simulate numerically in large models.

The problem of direct simulation was avoided in an early study by
\cite{Beasley02} who used a semi-analytic model of galaxy formation in
a cold dark matter (CDM) Universe. They assumed that globular clusters
would form in each galaxy in numbers proportional to the numbers of
stars forming (given by the semi-analytic model). They could then
ascribe chemical properties to the GCs according to the same
model. Their model successfully reproduce bimodal GC populations and
other observations, but only if they invoke a truncation of
metal-poor GC formation at redshift $z>5$. This truncation was also
investigated by \cite{Bekki05} in a collisionless dark matter
simulation of the formation of a single galaxy (mass resolution $4
\times 10^7 M_\odot$) with the simplifying assumption that GCs formed
in all low-mass subhaloes forming before some truncation
redshift. Bekki shows that the final ($z=1$) radial distribution of
the objects is very sensitive to the truncation redshift which was
set at $z\approx 15$ to match the GC distribution of the Milky Way. In
a more general study of structure formation \cite{Moore06} identified
reionisation as the process responsible for the truncation, assuming
that it took place by $z \approx 12$. They went on to suggest that the
radial distribution of GCs and satellite galaxies could be used to
constrain models of the reionisation process. These last two studies
established the importance of reionisation and the truncation of
globular cluster formation, but were not able to calculate the
important parameters directly.

A more comprehensive approach to the problem was taken by \cite{Kravtsov05} who
combined both gas and N-body codes to model a Milky Way sized galaxy, although
the simulation only ran to a redshift of $z=3$. Their model used the assumption
that GCs formed in sufficiently massive giant molecular clouds---themselves in
the disks of protogalaxies. This work was successful in matching several
observed GC properties such as masses and sizes, but could not predict
present-day positions. It was---like the earlier work---very reliant on
assumptions about the conditions necessary for GC formation.

The N-body (dark matter) and semi-analytic approaches have recently been
combined in a large scale cosmological simulation by \cite{Bekki08} to model the
dynamical and chemical properties of GCs in a wide range of galaxies. The
simulation covers a large volume, so the mass resolution is relatively low
($\sim 3 \times 10^8 M_\odot$) compared to GC masses: they simulate GC formation
by assuming every virialised dark matter halo of 10 or more particles will form
a GC. They also rely on a truncation of metal-poor GC formation due to
reionisation at $z_{trunc} = 6$ (this value based on quasar data; see discussion
below). From their model \cite{Bekki08} obtain old ages for both metal-rich
GCs (peaking at z$\sim 4$) and metal-poor GCs (peaking at z $\sim 6.5$) and also
obtain bimodal metallicity distributions for about half the galaxies. They also find the
galaxies without bimodal distributions tend to be small galaxies which lack
metal-rich GCs. The model also produces more centrally concentrated
distributions for the metal-rich GCs than for the metal-poor GCs although
physical origin of this difference is not specifically discussed.


A common feature of the simulations described above is the reliance on \textit{ad hoc}
assumptions to identify the sites where GCs form. This was inevitable because
the mass resolution was too low to resolve individual GCs. In this paper we
present the first study of globular cluster formation based on a simulation in
which globular cluster masses are well resolved.

We use the Aquarius suite of simulations \citep{Springel08}, the
highest resolution simulations of Milky Way sized haloes done to
date. It must be emphasised that this is the first work in which each
GC formation site is \textit{directly} resolved with a
\textit{minimum} of about 2000 particles. Although this first paper is
based strictly on the dark matter components of the simulation, the
exquisite resolution allows us to calculate the conditions for
metal-poor GC formation directly. We also include a more qualitative
model for the formation of metal-rich GCs which successfully predicts
their centrally concentrated distributions.

Our model for metal-poor globular cluster formation is different from
previous work in two major respects. First, we model the formation by
directly identifying when and where the early haloes first reach a
temperature of $10^4$K, the threshold temperature above which rapid
cooling takes place leading to collapse and star formation.  Secondly,
we do not assume an arbitrary value for the redshift when globular
cluster formation is truncated. Instead, we directly estimate the
number of ionising photons emitted by these early clusters and stop
their formation when there are enough photons to ionise the entire
galaxy.

We have used the simulation to test two models for the formation of the
metal-rich GC population: (i) the stripping of GCs from disrupted satellite
galaxies and (ii) the formation of star clusters in the gas disk of the forming
galaxy, triggered by large merging events. We find that only the second model
can produce the centralised distribution of metal-rich GCs as observed in the
Milky Way.

The structure of this paper is as follows: In Section
\ref{sec:method}, we describe the Aquarius suite of cosmological
simulations and our models for forming globular clusters in the
simulations. In Sections~\ref{sec:mp} and \ref{sec:mr}, we present the
results of our models in detail, notably comparing the spatial
distributions with observations, first for the metal-poor GCs and then
for the metal rich GCs. We then discuss our results in
Section \ref{sec:disc} and conclude by summarising up our results
and future work in Section \ref{sec:conc}.

\section[model]{Models of Globular Cluster Formation}
\label{sec:method}

In this section we describe the methodology of our study: how we
identify the globular cluster formation sites in the Aquarius
simulation data. We start by summarising the relevant details of the
simulations themselves. We then describe our models for metal-poor and
metal-rich globular cluster formation in detail in Sections
\ref{sec:methodmp} and \ref{sec:methodmr}, respectively.

\subsection{The Aquarius Simulations}
\label{sec:aquarius}

The Aquarius Project actually consists of simulations of six different
Milky Way-sized galaxies, one of which is analysed in this
paper. Although a detailed description of the simulations can be found
in \citet{Springel08}; hereafter S08, we review the pertinent details
here.

The starting cosmological parameters for Aquarius are the same as used
in the Millennium Simulation \citep{Springel05} project which are
consistent with both the WMAP1 and WMAP5 results (\citealt{Bennett03}; \citealt{Spergel07}). Halo formation is
tracked within a periodic box of side 100 $h^{-1}$ Mpc in a cosmology with parameters $\Omega_m$ = 0.25, $\Omega_\Lambda$ =
0.75, $\sigma_8$ = 0.9, $n_s$ = 1, and Hubble constant $H_0$ = 100 $h$
km\ s$^{-1}$\ Mpc$^{-1}$ = 73 km\ s$^{-1}$\ Mpc$^{-1}$. We use a baryon fraction  $\Omega_b/\Omega_m=0.18$ to convert from dark matter mass to baryonic mass.

Millennium simulation haloes
of roughly Milky Way mass and without close neighbours at $z = 0$ were
selected for resimulation using $900^3$ particles in a box of dimension 10 $h^{-1}$ Mpc.
Identifying the Lagrangian region from when each halo
formed, the mass distribution was rerun at a much higher spatial and
mass resolution. Distant regions were sampled with more massive
particles, but retained sufficient resolution to ensure an accurate
representation of the tidal field at all times. For greater detail on the simulation, see S08.

A major feature of the simulations is the identification of
substructure---the bound mass concentrations that will grow and merge
over time to build structure. These are identified and measured in
Aquarius using the same {\sc subfind} \citep{Springel01} algorithm
used for the Millennium simulation. Although, {\sc subfind} outputs are commonly called `subhaloes', we use the words subhalo and halo interchangeably to represent potential GC structures. This is because a potential GC can be classed as either an object within much larger parent object (i.e. a subhalo) or an object outside the virial radius of any other halo (i.e. a halo). For future reference, wherever we state the word `halo' and `subhalo', we are referring to the same thing; the {\sc subfind} outputs.

The way in which the haloes are
linked between time steps works as follows. For each halo we take the most bound 10$\%$ of its particles and determine
which halo they belong to at the next snapshot. The halo at the
later time with the largest number of these particles is identified as the
descendant. In most cases this is all that is required. However, there are occasional cases where subfind fails to identify a
halo at one or more snapshots but picks it up again at a later time.
For example, this can happen if a halo passes close to the centre of a
larger halo. We attempt to deal with this by looking for descendants more
than one snapshot later if a halo is not the most massive progenitor of
the descendant identified using the procedure described above or if no
descendant is identified. A halo which exists up to 3 snapshots later
will be identified as the descendant if it contains more than half of the
10$\%$ most bound particles from the original halo and has no identified
progenitors. 

All these different data structures between time steps for each simulation are stored in a
database very similar to that of the Millennium
project\footnote{http://www.g-vo.org/Millennium}. We also
make use of the raw particle data in cases where we are unable to obtain the {\sc subfind} halo information to track material from haloes that have merged. This is discussed in further detail in Section \ref{Sec:disrupted}.

Each of the six Aquarius haloes was calculated using at least two
different particle masses (`resolutions') to test for
convergence. We selected our highest resolution halo (`A'  halo) for our GC study in
this paper and analysed it at two different
resolutions so as to check our results for any dependence on
simulation resolution. A summary of the two data sets 
is given in Table \ref{tab:sim}. 

\begin{center}
\begin{table*}
\caption{The basic parameters of the Aquarius simulation data used in this paper.
}
\label{tab:sim}
\begin{minipage}{165mm}
 \begin{tabular}{cccccccccc}
\hline
Name & \textbf{$m_\mathrm{p}$} & $\epsilon$ & $N_{hr}$ & $N_{lr}$ &
$M_{200}$ & $r_{200}$ & $M_{50}$ & $r_{50}$ & $N_{50}$\\
& $[M_\odot]$ &[pc]& & & $[M_\odot]$ & [kpc] & $[M_\odot]$ & [kpc] & \\
\hline
Aq-A2 & $1.370 \times 10^4$ & 65.8 & 531,570,000 & 75,296,170 & $1.842 \times 10^{12}$ & 245.88 & $2.524 \times 10^{12}$ & 433.52 & 184,243,536 \\
Aq-A3 & $4.911 \times 10^4$ & 120.5 & 148,285,000 & 20,035,279 & $1.836 \times 10^{12}$ & 245.64 & $2.524 \times 10^{12}$ & 433.50 & 51,391,468 \\
\hline
\end{tabular}

{Notes: $m_p$ is the particle mass, $\epsilon$ is the
  Plummer equivalent gravitational softening length, $N_{hr}$ is the number of
  high resolution particles, and $N_{lr}$ the number of low resolution particles
  filling the rest of the volume. $M_{200}$ is the virial mass of the halo,
  defined as the mass enclosed in a sphere with mean density 200 times the
  critical value. $r_{200}$ gives the corresponding virial radius. We also give
  the mass and radius for a sphere of overdensity 50 times the critical density,
  denoted as $M_{50}$ and $r_{50}$. Note that this radius encloses a mean
  density 200 times the background density. Finally, $N_{50}$ gives the number
  of simulation particles within $r_{50}$.}
\end{minipage}

\end{table*}
\end{center}

\subsection{Formation of Metal-Poor Globular Clusters}
\label{sec:methodmp}

\subsubsection{Temperature Threshold}

We use a relatively simple model to identify where the metal-poor GC-type
objects would form within the Aquarius simulation based on the conditions
necessary for the collapse of a proto-GC gas cloud and subsequent star
formation. Recent simulations all rely on a similar model for GC
formation (dating back to \citealt{Peebles84}), but the resolution
of Aquarius allows us to measure the main parameter -- temperature -- directly.

The gas clouds cannot collapse without an efficient cooling mechanism:
in the absence of significant amounts of heavy elements, the main
cooling processes are the collisional excitation of hydrogen and
helium, radiative recombination of hydrogen, and bremsstrahlung
\citep[e.g.][]{Nishi02}. The typical cooling function for primordial
gas in the equilibrium state reveals an extremely rapid increase in
cooling rate as the temperature rises through 10$^4$K. We therefore
adopt 10$^4$K as a temperature threshold, above which gas clouds can
efficiently cool and collapse to form GCs.

Given the very large mass of gas required to form a globular cluster
(see Section~\ref{sec:mpreion}), the early proto-cluster gas clouds
must form in the potential wells of the dark matter subhaloes
identified by {\sc subfind}.  By assuming the gas is in quasi-static
equilibrium with the dark matter, we can use the virial theorem to
relate the 1-D internal velocity dispersion measured of the dark
matter subhaloes ($\sigma_v$) to the (inferred) virial temperature of
the gas, $T_v$:
\begin{center}
\begin{equation}
\sigma_v^2=\frac{kT_v}{\mu m_{\rm H}},
\label{virialtemp}
\end{equation}
\end{center}
where \changed{$m_\mathrm{H}$} is the mass of a hydrogen atom and
$\mu$ is the mean molecular weight of the gas. We adopt molecular
weight of $\mu = 0.58$, appropriate for a fully-ionised, primordial
gas. Our 10$^4$\ K temperature threshold therefore corresponds to a
1-D velocity dispersion of $\sim$ 11.9 kms$\1$.

We identify metal-poor GC formation sites by searching the entire
merger tree of the simulation for any subhaloes that exceed the $10^4$
K threshold for the first time. 

\subsubsection{The Final Positions of Disrupted Subhaloes}
\label{Sec:disrupted}
Although we have excellent resolution to directly locate where
candidates first form, it is no longer possible to locate these
structures as distinct subhaloes
at redshift zero if they have undergone merging events with either
each other or the central host halo since it leads to their complete
destruction.  We address this problem by using the most bound particle
of each subhalo as a marker to track its final position. This is 
equivalent to assuming that the collapsed baryonic gas forming a GC at
the centre of each subhalo will follow a similar trajectory to the
most-bound particle.  We simply search the final simulation snapshot
at $z = 0$ to locate where each of these uniquely identified particles
reside. This directly allows us to follow GC formation sites through
to the present day.

As we show below, the majority of the haloes containing GC formation
sites merge with the central halo by the present day, but some survive
in separate haloes at a range of distances. We include both groups in
our model (both merged and surviving haloes) but we restrict our
discussion to the GC candidates that end up associated\footnote{By associated we
mean either fully merged with the main halo or in surviving haloes that
are within two times the half-mass radius (2\,$r_{1/2}=150$\,kpc) of
the main halo at redshift zero.} with the main
Milky Way halo in the simulation at redshift zero. There are a few objects in the outer
halo that do not satisfy this condition: we discuss these further in
Section~\ref{sec:disc}.

\subsubsection{Reionisation within Aquarius from the first GCs}
\label{sec:methodreion}

Previous simulations of GC formation have found it necessary to truncate the
formation process after a certain redshift in order to avoid producing
unrealistically large numbers of clusters (\citealt{Kravtsov05}, \citealt{Bekki05,Bekki08}). The truncation redshift has generally
been set on the basis of external estimates of the redshift of
reionisation. In this paper we take a different approach, based on an internal
calculation of the ionisation from the first star clusters themselves.

Current estimates of when the Universe as a whole was reionised are based on
absorption line studies of high-redshift quasars. Observations of Ly$\alpha$
systems in high-redshift quasars indicate the inter-galactic medium (IGM) was
fully ionised at $z \sim 6$ (\citealt{Gnedin02}). The relative scarcity of
quasars at redshifts greater than $z \sim 4$, points to another source of
ionisation. Unless far more quasars are found, the photoionising contributions
of high-$z$ massive stars seem to be the only plausible reason to account for
the missing radiation. The key parameter of measuring their contribution
observationally (from the quasar spectra) is the escape fraction of Ly$\alpha$
photons from galaxies into the IGM, $f_{\rm esc}$. Unfortunately this
approach can not yet be applied: there is a wide range of estimates of
$f_{\rm esc}$ (\citealt{Madau96,Bianchi01,Ricotti02}) and at best they
are upper limits, as the absorption of ionising radiation from the molecular
cloud and dust extinction are ignored.

In this work we estimate the ionising contribution of massive stars within the
forming galaxy directly from the simulation. We assume that the local ionising
radiation is dominated by flux from the first GCs to form. We calculate the
number of ionising photons from each GC using a Salpeter initial mass function
(IMF) based on Population II star formation models from the Starburst99 code by
\cite{Leitherer99}. We do not aim to ionise the entire IGM; instead we
specifically ask if local ionising photons from the first GCs are sufficient to
ionise the \textit{intra}-galactic medium of the forming galaxy by some
redshift, $z_{ion}$. Our full calculation in Section \ref{sec:mpreion} shows
that this does happen, and at earlier redshifts than given by the quasar
estimates.

\subsection{Formation of Metal-Rich Globular Clusters}
\label{sec:methodmr}

It is more difficult to model the formation of the metal-rich GC population in
our simulation. First, the defining chemical make-up of this population can only
be modelled if we include baryonic gas processes. Secondly, although these
clusters are slightly younger than the metal-poor GCs, they are actually the
more centrally concentrated of the two populations in the Milky Way. This is
counter to what happens in hierarchical merging processes where the first
objects to form end up most centrally concentrated. We have considered two very
simple models that can be tested qualitatively with our current simulation data.

\subsubsection{Model 1: Tidal stripping of satellite dwarf galaxies}
\label{sec:methodtidalstripping}
Based on previous models of tidal interactions removing the outer
envelopes of dwarf elliptical (dE) galaxies (\citealt{Bekki05}), we
investigated a stripping model of GC formation. The basic idea here is
that the dwarf galaxies have formed their own (metal-rich) GCs that
survive to join the main halo after the dwarf is disrupted. We tested
this model by identifying as ``stripped'' any halo which merges with a
more massive halo (typically, the mass increase in such a merger is a
factor of 10 or more).  The maximum mass of the progenitor halo,
before merging, is taken to be proportional to the mass of the nuclear
globular cluster that forms within it.  

As we show below in Section~\ref{sec:mr}, this results
in far too extended a distribution. This is the problem mentioned
earlier that, in hierarchical models, younger objects have less
centrally-concentrated distributions than older ones.

\subsubsection{Model 2: Major mergers with the central halo}
\label{sec:methodmajormergers}
The proposal that metal-rich GCs are formed in gas-rich mergers of interacting
galaxies has been around for quite some time. \cite{Toomre72} were the first to
investigate these merger events in detail but it was not until much later that
the formation of GCs in this framework was discussed \cite{Schweizer87}. {\em
  Hubble Space Telescope} observations of young massive star clusters forming in
the merging Antennae galaxies (\citealt{Whitmore95}) provided strong motivation
for this model (see also \citealt{Holtzman92}) which remains the subject of many
studies (\citealt{Zepf06}).

We adopt the following simple model of metal-rich GC formation by gas-rich
mergers. We search the merger tree for any halo (above some mass threshold)
which merges with the central halo. We adopt the
premise that during the merger event, stars will form via perturbations in and
around the gas disk of the central halo at that particular redshift. Since the
merging haloes are destroyed in this process, the location (radius with respect
to the central halo) of where the stars form can be approximated by the radius
of the gas disk. This we assume is a fraction of the half-mass radius of the
central halo at the redshift of the merger event. Since we have access to the
half-mass radius of the central halo at each time step, we essentially count the
number of merger events at a given time step, and say that all of those
in-falling haloes will create stars at some fixed fraction of the half-mass
radius.

We adopt the major-merger scenario here because it is motivated not
only by a rich range of observational results, but it is the only
model we can find which can make metal-rich GCs with a more
concentrated distribution than the metal-poor GCs, as is observed in
the Milky Way.

\section{Metal-Poor Globular Clusters}
\label{sec:mp}

\begin{figure}
\centering
\epsfig{file=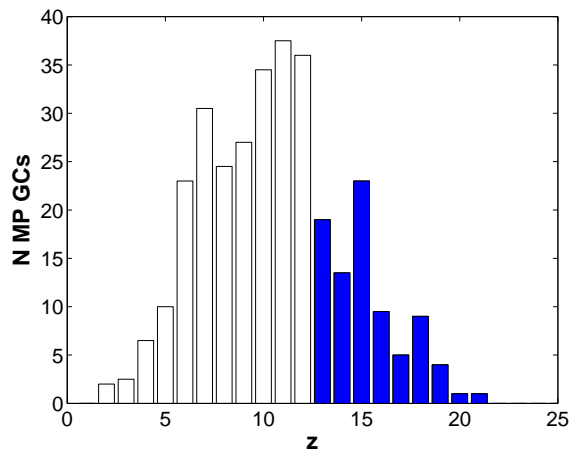,width=8.4cm}
\caption{Formation history of all metal-poor (MP) GCs. The figure shows the
  redshift when they first exceed a virial temperature of $10^4K$.
  The filled bars represent those haloes that form before $z_{ion}$
  ($z_{ion}$ = 13). The hollow bars are the remaining haloes identified
  above the temperature threshold but are suppressed due to the
  ionising contributions (see Sec. ~\ref{sec:mpreion}) from the (blue)
  haloes that formed earlier. By z = 13, the entire Galaxy has been
  ionised.}
\label{fig:NvsZALL}
\end{figure}

We identify candidate haloes for metal-poor globular clusters using the
method described in Section~\ref{sec:methodmp}.   The redshift
distribution of the haloes is shown in Figure~\ref{fig:NvsZALL}.

There are two things to note about this distribution.  Firstly, there
are far too many GC candidates for a Milky-Way sized galaxy, and
secondly the distribution extends down to low redshift.  We
address these concerns by estimating, in the following section, the
redshift at which the globular clusters that form can reionise the
protogalaxy.  Those haloes that form after reionisation are shown as
hollow bars in the figure.

\subsection{Reionisation Contributions}
\label{sec:mpreion}

We define an ionising efficiency, $F_{\rm ion}$, to determine the mass
of gas ionised by each globular cluster in our model. $F_{\rm ion}$ is
the mass of baryons ionised by the GC divided by the baryonic mass of
the halo in which the GC first forms (assuming each of our GC
formation sites forms only one GC). Each globular cluster can
therefore ionise a region greater in baryonic mass than itself by a
factor of
\begin{equation}
F_{\rm ion}=f_{\rm sfe}\,\bar{q}\,f_{\rm esc},
\end{equation}
where $f_{\rm sfe}$ is the star-formation efficiency (the fraction of the
protocluster baryonic mass that forms  stars), $\bar{q}$ is the mean number of ionising
photons per baryon locked up in stars, and $f_{\rm esc}$ is the fraction of
those photons that escapes the cluster.  In writing down this expression we
assume that one ionising photon per baryon is sufficient to ionise the
surrounding gas out to $2 r_{1/2}$ (where $r_{1/2} = 75$ kpc is the dark matter half-mass radius of the main halo in both simulations). This radius contains the majority of the GCs and of the baryonic mass of the galaxy.  This is a reasonable
approximation given the uncertainties in the other factors.

For a self-gravitating gas cloud, star-formation efficiencies of order one third
or more are required in order to form a bound cluster
\citep{Baumgardt07}.  However if,  if not all the gas in a halo ends up
in the proto-cluster the SFE could be considerably lower.  Here we
take $f_{\rm sfe}=0.07$ because that gives masses for the GCs
  that seem to agree with observations - see Section \ref{sec:masses}. 

The number of ionising photons emitted per baryon of stellar material depends
strongly upon the stellar mass.  We use the the Population~II efficiency curve of
Fig. 2 from \cite{Tumlinson04} (calculated using the STARBURST99 code of
\citealt{Leitherer99}).  Averaging over a Salpeter IMF (see
Appendix) gives $\bar{q}\approx10\,000$.  This figure
could be raised by moving towards a more top-heavy IMF.  Note that the
lifetime of a 10\,$M_\odot$ star is around 40 million years, of order
the dynamical time of the first globular cluster haloes that form and
less than the dynamical time in later ones.  It is a reasonable
approximation, therefore, to assume instantaneous feedback.

As discussed in Section~\ref{sec:methodreion}, the escape fraction of
photons is the most uncertain factor.  To obtain the correct
  number density of metal-poor GCs, we require $f_{\rm esc}=0.3$,
which is not unreasonable given the escape fraction in the early
Universe would be considerably higher than we observe locally.

\begin{figure}
\centering
\epsfig{file=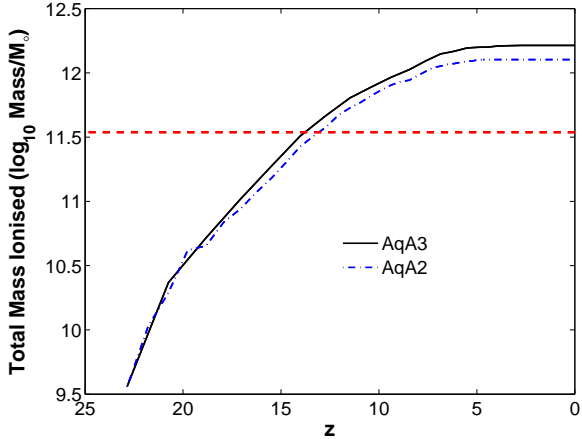,width=8.4cm}
\caption{Cumulative mass of baryons ionised by metal-poor GCs for both high
  (black, solid line) and low (blue, dash-dot line) resolution runs. The total
  baryonic mass of the central galaxy at the present day is given by the red
  dashed line.}
\label{fig:ionised}
\end{figure}

Putting all these factors together, we have $F_{\rm ion}\approx 210$.  Figure~\ref{fig:ionised} shows the cumulative
mass of ionised gas for the AqA2 halo, starting at high redshift and
moving towards the present.  Note that we only include the contribution from GCs that end up within $2 r_{half} = 150$ kpc of the central halo in the present day. The two curves correspond to the two
different numerical resolutions and show good agreement -- at
  $z=13$ the difference between the two curves represents less than one
  output time in the merger tree construction.  The dashed line shows
the total baryonic mass of the halo at the present day (i.e.~the mass
within a sphere centred on the most-bound particle and enclosing a
mean density of 200 relative to the critical density).

The radiation produced by the
metal-poor globular clusters is sufficient to ionise a mass equal to
that of the present-day galaxy by a redshift of about 13 (Fig. \ref{fig:ionised}).  There is a
lot of uncertainty in this estimate and varying $F_{\rm ion}$ by
a factor of two could give an estimated ionisation redshift
in the range 10 to 15.  However, using $F_{\rm ion}$ = 210 (z = 13) gives a number of clusters that
agrees well with observations and so we stick with that for the rest
of our analysis.

Using this value for the reionisation cut in AqA2, we define a total sample of 173 metal-poor globular clusters, of which 125 lie within the 2$r_{1/2}$ of the central halo. The majority of these (105) are no longer in separate dark matter haloes, but have merged with the central halo. The number of GCs we form is quite consistent with the number of metal-poor GCs observed in the Milky Way (103 with [Fe/H] $<-1$; \citealt{Harris91}).

\subsection{Masses}
\label{sec:masses}
In order to calculate the present day masses of our
metal-poor GCs, we first need to know what fraction of the
  baryonic mass of a gaseous protocluster becomes locked up in stars.
\cite{Baumgardt07} and \cite{Weidner07} found that in order to form a
bound star cluster, a protocluster requires a star formation
efficiency of approximately 0.3. However, this can be much lower if
not all the gas within the halo cools to become part of the
protocluster: for example, the cluster may form from a small
fraction of the gas that has condensed at the centre of the halo.
  In this paper we adopt a value of 0.07 as this gives current-day GC
masses that seem to agree with those in the Milky Way. 

We then require an estimate of how much mass loss due to
stellar evolution (winds, SNe) and dynamical evolution from tidal
stripping and evaporation affects the cluster. \cite{Kruijssen08}
showed that a typical globular cluster loses $\sim$70$\%$ of its
initial stellar mass and that is the value that we adopt
  here.  Overall, this means we have from an initial dark matter mass
of M$\rm_{DM,0}$, a final present day stellar mass of
$0.18\times0.07\times0.3 \rm{M}_{DM,0} = 0.0038\ \rm{M}_{DM,0}$.

\begin{figure}
\centering
\epsfig{file=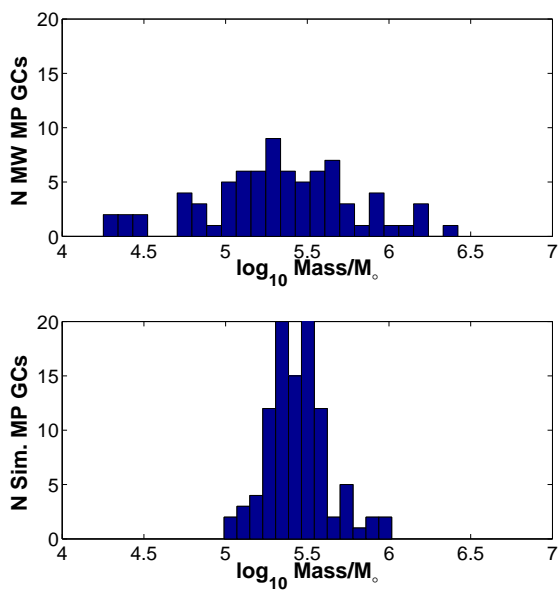,width=8.4cm}
\caption{Comparison of the masses of simulated metal-poor (MP) globular
  clusters with the observed masses of Galactic GCs. \textit{Top
    panel:} Number of Milky Way GCs with a given mass for those with
  available integrated V-band luminosities (calculated from the Harris
  et al. (2006) catalogue assuming $M/L_v$ = 3 and including only
  metal-poor GCs with [Fe/H]$<-1$). \textit{Bottom panel:}
  Calculated present day masses of our GC candidates (AqA2 resolution)
  that formed before $z = 13$.}
\label{fig:MWSIMmass}
\end{figure}

Figure \ref{fig:MWSIMmass} compares both our candidates identified
before $z = 13$ to those of the Milky Way with [Fe/H]
$<-1$ (assuming $M/L_v = 3$ for 13\,Gyr old stellar
population (\citealt{Maraston05}). We obtain a range of
masses between $\sim$ $10^5$ -- $10^6 \ \rm{M}_\odot$. The masses are
consistent with the mean of, but have a narrower spread than, the
observed Milky Way GCs mass distribution. This may be due to
observational errors in calculating the mass and/or scatter in the
mass-loss from individual clusters.

\subsection{Ages}
With the reionisation cut, all the metal-poor GCs form in the redshift
range 22-13, corresponding to look-back times of 13.3-13.5\,Gyr. The
precise age of the globular clusters presumably mimics the assembly
time of the galactic halo, but the prediction of a narrow spread in
ages will hold for all galaxies.

This is broadly consistent with observations of the Milky Way.  From
homogenous age dating of 55 Galactic GCs carried out by
\cite{Salaris02}, we know that the majority of the metal-poor GCs
([Fe/H] $<$ -1.6) formed approximately $12.2\pm0.6$ Gyr ago with a narrow
spread in ages, consistent within the errors with a single formation
redshift.

\subsection{Spatial Distribution}

\begin{figure}
\centering
\epsfig{file=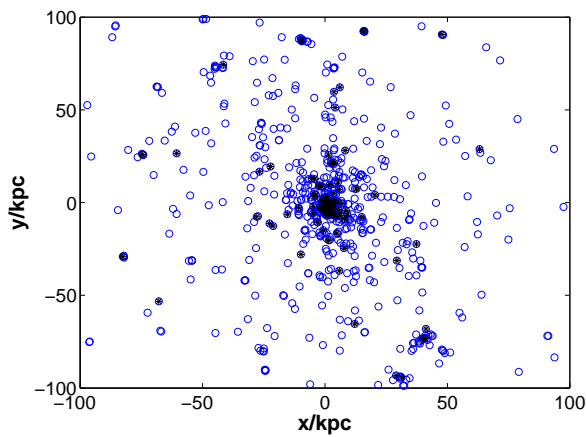,width=8.4cm}
\caption{Present day spatial x-y distribution demonstrating the effect
  of including (filled circles) and excluding (open circles) metal-poor GC
  truncation at z$_{trunc}$ = 13 for the AqA2 halo.  Without including
  reionisation, there are far too many GCs, particularly at large
  radii.}
\label{fig:spatialxy}
\end{figure}

Figure~\ref{fig:spatialxy} shows a projection of the final GC locations
relative to the most-bound particle in the host galaxy.  Filled
circles represent those clusters that form before our reionisation
cut, and open circles those that form afterwards.  The latter
population is much more extended than the former: this is to be expected as they form later.

\begin{figure}
\centering
\epsfig{file=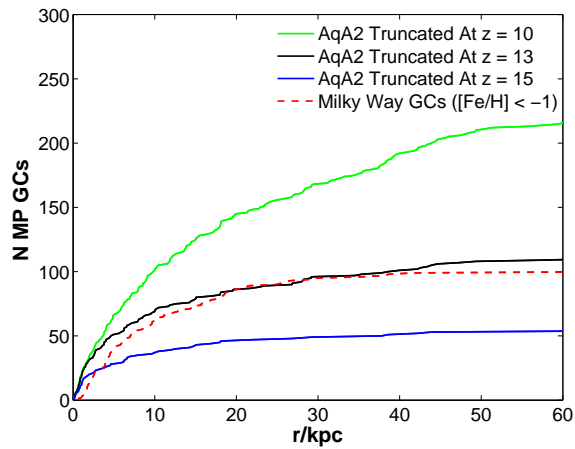,width=8.4cm}
\caption{Spatial distribution of both model and observed metal-poor
  globular clusters. The curves show the cumulative radial
  distributions of each population. Three different model populations
  are shown, corresponding to the three ionisation cuts discussed in
  the text (solid lines). The dashed line shows the observed
  distribution of metal-poor globular clusters in the Milky Way
  \citep{Harris91}.}
\label{fig:MPspatial}
\end{figure}

Fig. \ref{fig:MPspatial} shows a present day cumulative radial
distribution of the Milky Way metal-poor GCs and the AqA2 metal-poor
GCs using three different reionisation cuts, correspoding to
$F_\mathrm{ion}=$105, 210 and 420. The radial distribution of Milky
Way GCs and of our model GCs are in agreement, whereas earlier-forming
haloes are too centrally-concentrated and later-forming ones too
extended. This is very strong evidence that the metal-poor globular clusters
do indeed form in the high-redshift haloes that we have identified.

\subsection{Kinematics}
\label{sec:mpkinematics}

The limited observational evidence on the kinematics of metal-poor globular
cluster populations in late-type galaxies is described in \citet{Brodie06}.
There is some indication that different sub-populations may have different
rotation properties, as if the galaxies have been built up from mergers of
smaller systems.  This is consistent with the idea that metal-poor globular
clusters formed early on, before their host galaxy.

Overall, there appears to be little net rotation of the observed metal-poor GC
population.  Our models agree with this result, showing rotation speeds of order
10\,km\,s$^{-1}$ that are consistent with no net rotation within the sampling
errors.

The velocity dispersion of the model globular cluster population is
radially biased, having an aniotropy parameter of
$\beta={3\over2}(1-\sigma^2/\sigma_r^2)\approx0.6$ (here $\sigma_r$
and $\sigma$ are the root-mean-square radial and total velocities
relative to the galactic centre, respectively).  The observational
evidence is currently too weak to place strong constraints on $\beta$.

\section{Metal-Rich Globular Clusters}
\label{sec:mr}

\begin{figure}
\centering
\epsfig{file=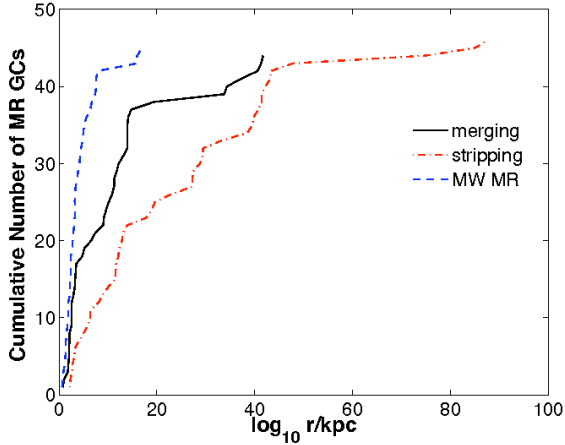,width=8.4cm}
\caption{The radial distribution of metal-rich (MR) GC candidates from the
  stripping model (dash dotted line) and the merging model (solid line)
  compared the their distribution in the Milky Way (dashed line).}
\label{fig:mrprof}
\end{figure}

The Milky Way metal-rich GC population is more centrally-concentrated
than is the metal-poor one.  This is a problem for any formation model
that invokes accretion from haloes more massive, and hence
later-forming, than those used to define the metal-poor population.
As an example, we show in Figure~\ref{fig:mrprof} the current-day
distribution of metal-rich GCs in a model in which they are the nuclei
of stripped dwarf galaxies that have been disrupted by the tidal
forces of the Galaxy.  In this plot, we estimate the current location
of the GC from that of the most bound particle in the dwarf galaxy at
the timestep before it lost its identity.  In order to match the
observed number of metal-rich GCs in the Milky Way, we have included
dwarf galaxy haloes whose maximum mass before disruption exceeded
$8\times10^8\Msun$ but other mass-cuts produce similar profiles.  As
can be seen from the plot, the radial distribution of such objects is
far too extended and cannot possibly represent the metal-rich GC
population.

To find a distribution that is more centrally-concentrated than that
of the metal-poor GCs we must abandon models that form clusters within
sub-haloes and instead form them directly within the galaxy itself.  As
mentioned in Section~\ref{sec:methodmajormergers}, there is
observational evidence for the formation of massive star clusters in
merging galaxies and so that is the model that we turn to here.

It is not obvious how large a merger is needed in order to generate
enough disturbance to trigger GC formation.  To match the number of
metal-rich GCs in the Milky Way, we assume that a single GC forms
whenever a galactic halo of mass $8\times 10^8\Msun$ or more merges
with a larger halo.  We further need to assume that the infalling halo
must have a mass of at least 1 per cent of the mass of the larger one;
otherwise large numbers of GCs would be formed by accretion of small
satellites onto the galaxy at late times.  

As the cold gas is located at the centre of the haloes, that is where
the GC will form.  For satellite haloes, we use the location of the
most-bound particle as a tracer of the GC position at later times.
For the main halo, we assume that the GC forms at a small fraction,
0.1, times the half-mass radius.

There are a lot of ad-hoc assumptions in this model: we don't know how
massive an infalling satellite must be to trigger GC production; we
don't know how many star clusters will form and what their mass will
be, and we don't know the precise location in which they will form.
Nevertheless, we present broad-brush results below in order to
demonstrate that the model GCs have approximately the right
properties and to motivate further study.

\subsection{Spatial Distribution}

The current-day spatial distribution of the GCs formed in the merging
model is shown as the dashed line in Figure~\ref{fig:mrprof}.  The
central concentration of GCs within 10\,kpc originates from mergers
with the main halo.  This agrees well with the observed distribution
of metal-rich GCs in the Milky Way.  In addition, about a
  quarter of the GCs result from mergers in satellite galaxies that
  later fell into the main halo. These latter objects have a more
  extended distribution and do not seem to have analogues in the Milky
  Way metal-rich GC population.

\subsection{Ages}
\begin{figure}
\centering
\epsfig{file=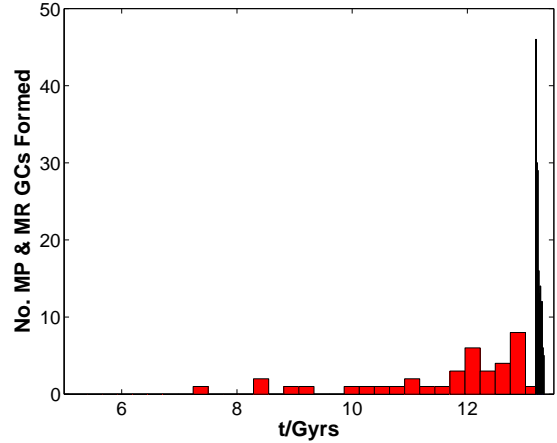,width=8.4cm}
\caption{The distribution of ages for metal-rich globular clusters
  (thick bars) formed using the merger model.  Also shown, for
  comparison, are the ages of the metal-poor globulars (thin bars).}
\label{fig:mrages}
\end{figure}

In Fig. \ref{fig:mrages}, we show the number of metal-rich GCs formed
as a function of look-back time.  Also shown, for comparison, are
the metal-poor GCs.  

We know from various observational studies of metal-rich GCs, that the
majority formed over a much larger temporal range than metal-poor GCs,
probably because they form via processes which require larger
dynamical times, \citep{DeAngeli05,Salaris02,Harris91}. These
observations are broadly consistent with our findings.  However, there
are seven GCs that have younger ages, between 0.8 and 3.8\,Gyr, which
we have not shown in Figure~\ref{fig:mrages}.  These do not have
analogues in the Milky Way, although we note that young ``intermediate
age'' globular clusters are found in M31 \citep[and references
  within]{SSL09}.  It is possible that they are an accident of the
formtion history of this particular galactic halo, or that the
galactic disk has become so depleted by these late times that it is no
longer susceptible to GC production in minor mergers.

\section{Discussion}
\label{sec:disc}

In this section we discuss more general aspects of our simulation of
the globular clusters starting with the achievements and
limitations of the models.

The model proposed in this paper for the formation of metal-poor GCs
is based on just two parameters: the temperature threshold for
cooling, and the ionising efficiency $F_{\rm ion}$. The first of these
is fixed by cooling physics and is not a variable in the model. The
second parameter, the ionising efficiency, is relatively uncertain
(especially for the escape fraction) as discussed in
Section~\ref{sec:methodreion}. In our calculation of $F_{\rm ion}$ we
have chosen reasonable values for the uncertain parameters, but they
were adjusted to give a good match to the total number of
metal-poor GCs observed in the real Milky Way.

Our adopted value ($F_{ion} = 210$) of the ionisation efficiency
corresponds to the suppression of GC formation after a redshift of
z$_{ion}$ = 13. (Even if we allow for a possible factor of 2
uncertainty in $F_{ion}$, this only gives a range of z$_{ion}$ =
10--15.) This redshift is significantly higher than the
estimates of the reionisation redshift of the universe derived from 
QSO studies (z$_{reion}$ = 6.4), but if GCs form high-mass stars first,
the local \textit{intra}-cluster medium is going to be far more
ionised than the rest of the Universe.
 
This model, whereby GC formation is truncated when the clusters
themselves reionise the remaining gas in the protogalaxy, has one
important feature: it naturally predicts that the number of metal-poor
GCs is proportional to the baryonic mass of the galaxy. For a
constant mass-to-light ratio this is directly equivalent to saying
that galaxies of this type will all have the same specific frequency
of globular clusters, $S_N$, as observed. The prediction of
proportionality is robust even with the uncertainty in the ionising
efficiency calculation; if that calculation becomes more precise we
will also be able to predict the absolute values of specific frequency.

Given that our model has involved some adjustment of one parameter to
match one observable (the total number of GCs formed), its success can
be demonstrated by testing it against other observations. In
Section~\ref{sec:mp} above we show that the model successfully
reproduces both the the radial distributions and the ages of the
metal-poor GCs. The agreement with these two independent measurements
gives us a high degree of confidence in our model. 

\begin{figure}
\centering
\epsfig{file=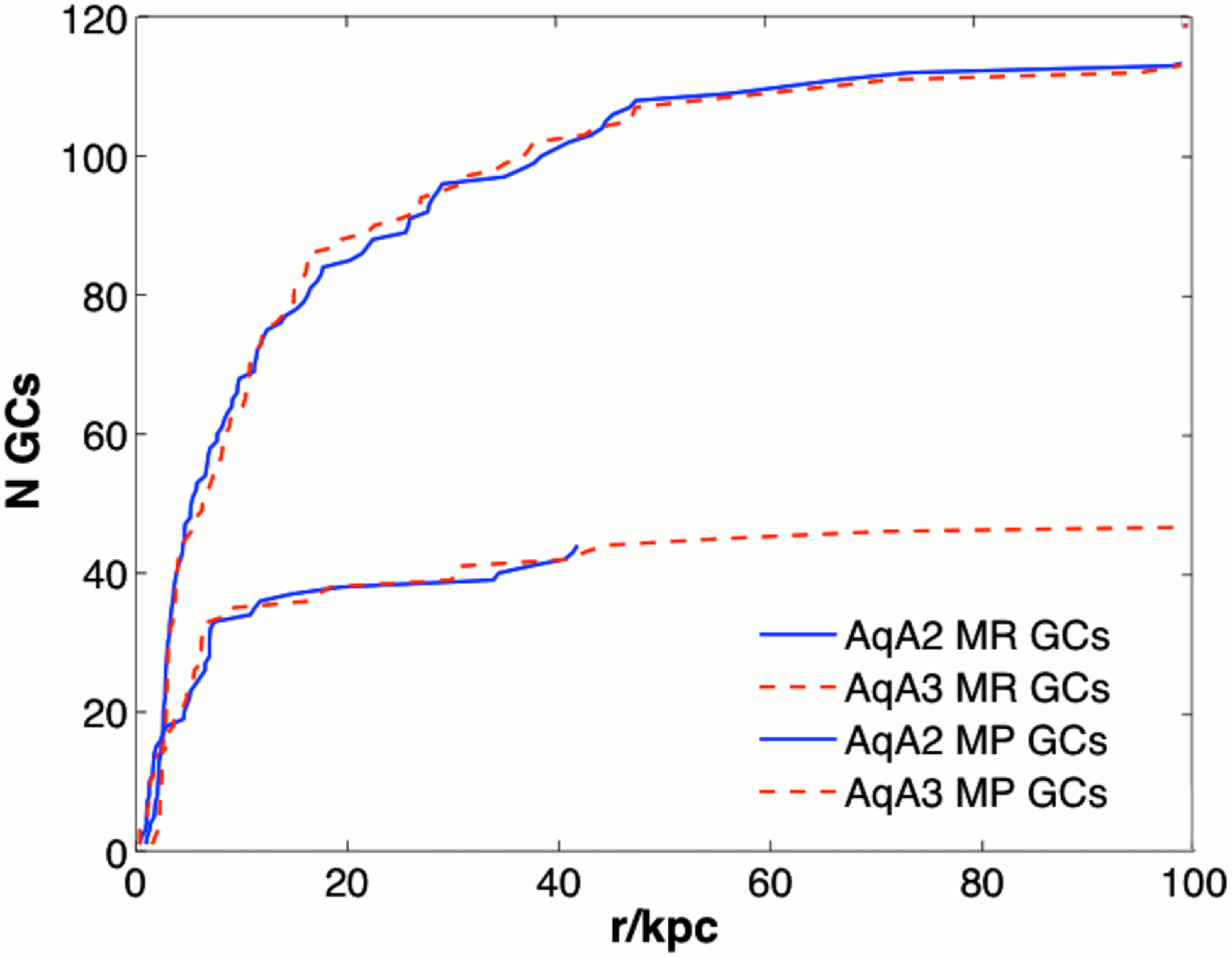,width=8.4cm}
\caption{Tests of the effect of the simulation mass resolution on our
  globular cluster formation models. We plot the radial distributions
  of both metal-poor and metal-rich GCs produced using both the
  original simulation (AqA2, $1.370 \times 10^4 M_\odot$, solid lines)
  and the lower-resolution AqA3 simulation
  ($4.911\times 10^4 M_\odot$, dashed lines). The metal-poor GCs
  are shown by the upper two curves and the smaller metal-rich GCs by
  the lower two curves. In
  both cases the results at the two resolutions are entirely
  consistent, showing that our results are not affected by the
  resolution of the simulations. 
  }
\label{fig:MPRconvergence}
\end{figure}

An important concern for both the metal-poor and the metal-rich models
is that the results should not be biased by the resolution of the
particular simulation used. We can test this in a very straightforward
manner for the current models by repeating the models with a
lower-resolution simulation: if the same results are obtained this
demonstrates that our models are not affected by mass resolution.  Our
main analysis uses the AqA2 simulation (mass resolution of $1.370
\times 10^4 M_\odot$); we have repeated both models with the
lower-resolution AqA3 simulation ($4.911\times 10^4 M_\odot$). In
Figure~\ref{fig:MPRconvergence} we show the result of the comparison
by plotting the radial distributions of both metal-poor and metal-rich
GCs produced in both simulations. For both models the agreement is
excellent: the total numbers produced agree to within 5 per cent and
a Kolmogorov--Smirnov test shows they are drawn from the same distribution. This agreement shows that
the results are not biased by the resolution of the simulations used.
Note that complete agreement is not expected for a variety of reasons:
the extra substructure will cause haloes to form at slightly different
times, and the trajectories of the most-bound particles will also be
altered. (The lower-resolution AqA3 model of metal-rich GCs extends to
larger radii in Figure~\ref{fig:MPRconvergence} than the corresponding
AqA2 objects only because of two haloes at large radii are just below
the mass limit in AqA2, but have slightly higher masses in the
low-resolution simulation.)

In our model of the metal-poor GCs we use the most bound particle (at the time of formation) of the dark matter halo hosting each GC to trace the final positions of GCs whose haloes have merged with the central galaxy halo. This may not be a good estimate of the final position if the GCs evolve in a different way to the single dark matter particle; possible processes are dynamical friction, tidal disruption and disk shocking. The time scale for dynamical friction (\citealt{Chandrasekhar43}) is inversely proportional to mass, so the GCs will be more affected by this process than the dark matter particles (which are of order 100 times less massive). For GCs of mass a few times $10^5 M_\odot$ at radii of 10 kpc the dynamical friction time scale is $10^{12}$ years, so this is unlikely to affect our results. Tidal effects can totally disrupt globular clusters on time scales of $10^{8}$ to $10^{11}$ years \citep{Kruijssen08}, so this may be an issue, but a much more detailed model of the evolution of the cluster candidates would be required to investigate this effect which we defer to future work. Finally, disk shocking is known to have a disruptive effect on GCs that pass through the disk of a galaxy. This has a time scale of about $6 \times 10^{9}$ years for Milky Way GCs (\citealt{Ostriker72}); this will affect the GCs in our model, but as it primarily removes loosely-bound stars from the outer parts of the GCs, it is unlikely to cause any separation of the most-bound particle from the GC.

We have also used the model to estimate the final masses of the
GCs. This involves a much higher degree of uncertainty as we have to
estimate the efficiencies of both cluster formation and their
subsequent evolution. We adjusted the star formation efficiency to
give mean masses consistent with Milky Way GCs, but the results then
suggest a slightly narrower range of mass than in observed
clusters. Interestingly, our results do not provide strong evidence
for a power law distribution in the masses of the metal-poor GCs at
formation as has been assumed in some studies of their subsequent
evolution \citep{Prieto08}. In future work, it would be valuable to
investigate the evolution of the GCs formed in our model in a similar
way.

\begin{figure}
\centering
\epsfig{file=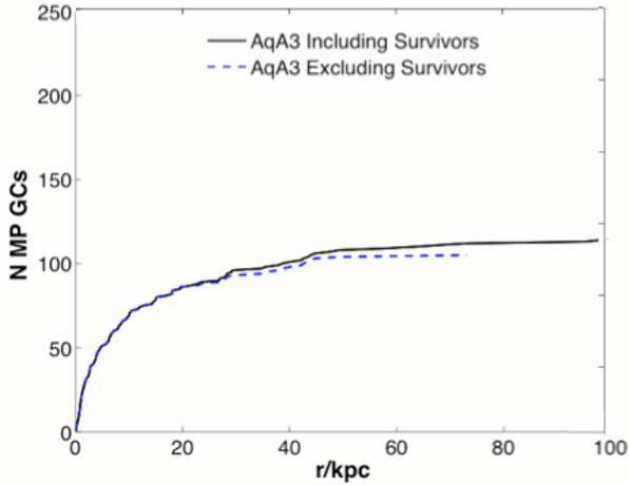,width=8.4cm}
\caption{The distribution of metal-poor GC candidates with surviving
  dark matter haloes. The plot compares the radial distribution of all
  metal-poor GC candidates with those that have
  merged (upper solid line)  completely with the central halo by redshift $z=0$. The lower dashed line shows the distribution of metal-poor GCs but excludes those which haven't merged with the central halo by $z=0$. The small
  gap between the two distributions indicates how few GCs have
  surviving dark matter haloes this close to the Galaxy.
}
\label{fig:survivors}
\end{figure}

The approach used above to estimate the GC masses was necessary in part because most of the GCs we analyse do not survive as separate haloes to the present day, but merge with the main halo. This is demonstrated by Fig.~\ref{fig:survivors}, which compares the total radial distribution at redshift zero of all the metal-poor GCs to those which have merged with the central halo. The gap between the two curves in the figure indicates the small number of GCs still found in surviving haloes: this is a very small fraction, less than 10 per cent within a radius of 100 kpc. The presence of these surviving haloes does suggest that some GCs associated with the Milky Way will have retained some dark matter, but at these relatively large radii these are may be associated with dwarf satellite galaxies rather than isolated GCs.

\begin{figure}
\centering
\epsfig{file=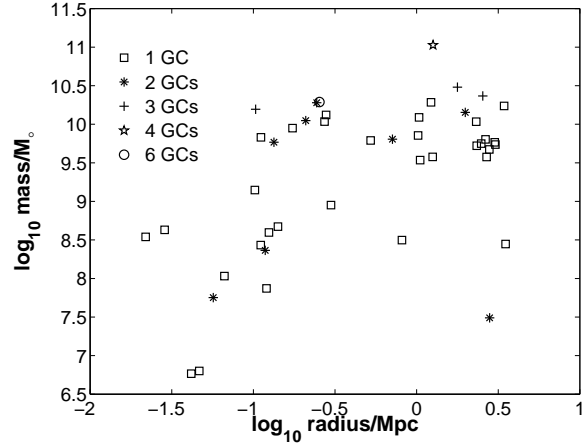,width=8.4cm}
\caption{The properties of haloes which have not merged with the
  central galaxy halo at redshift $z=0$ and contain metal-poor GC
  candidates. Each point gives the mass and projected radius from the
  central galaxy of a dark matter subhalo containing at least one
  metal-poor GC candidate. The symbols indicate how many GCs each
  contains (1: squares; 2: asterisks; 3: crosses; 4: stars; 6:
  circles). The distribution of masses and radii is very similar to
  that of observed dwarf galaxies in the Local Group.  }
\label{fig:survivormass}
\end{figure}

If we now consider all the GCs found in surviving dark matter haloes
distinct from the central halo in the full simulation (i.e.\ at larger distances from the main halo), we can measure their (dark matter)
masses and positions as shown in Fig.~\ref{fig:survivormass}. In this
figure there are 47 distinct haloes containing 68 GCs and the haloes
which contain more than one GC are indicated by different symbols. The
overall spread of properties is quite similar to that of dwarf
galaxies in the Local Group, both in distance and mass
\citep{Mateo98}. The objects in the figure can be divided into two
groups. The smaller
haloes ($<2\times10^9$\Msun) are nearly all at small radii ($<$100 kpc)
and are single GCs: these presumably correspond to small satellite
galaxies of the Milky Way that still retain some dark matter. On the
other hand, the more massive haloes ($>2\times10^9$\Msun) tend to lie at
large radii ($>$100 kpc) and often contain multiple GCs. These
correspond to the larger dwarf galaxies of the local group. We must
note, however, that our metal-poor GC formation model does not
necessarily apply to the more distant objects as they would not have
experienced the same ionisation environment as those forming closer to the
main halo. 

In a related study of the Aquarius simulations, \cite{Gao09} have investigated the formation of the very first stars,
systems with temperatures around $10^3$ K that form at redshifts of
$z=20$ or higher. In that context they also considered systems formed
by line cooling ($10^4$ K) at later times. These objects are very
similar to what we identify as candidate GC formation sites in this
paper, although Gao et al.\ refer to them as `first galaxies' and
focus on a comparison of the surviving objects with observed dwarf
galaxies around the Milky Way, obtaining similar results to those we
describe here.

As we note above, our model for the formation of the metal-rich GCs
relies on several assumptions: it is mainly intended to let us
determine if any class of model can come close to reproducing the
extremely concentrated distribution of these objects observed in the
Milky Way. Our merger model is significantly better than any
hierarchical model we have considered in reproducing the observed
combination of a centrally-concentrated distribution and young ages
for these objects. The model does produce a few metal-rich GCs at
larger radii and younger ages than observed in the Milky Way, but this
is not significant given the many other uncertainties in the model.

\section{Summary $\&$ Future Work}
\label{sec:conc}

In this study, we have made use of the exquisite resolution of the
Aquarius simulations to test plausible metal-poor and metal-rich GC
formation models. Here we summarise our main results and indicate our
plans for future work.

\subsection{Metal Poor GCs}

We adopted a relatively simple formation scenario for metal-poor GCs
by identifying all haloes which go above the 10$^4$K threshold required
to cool and form stars. We then measured the ionising contribution
from the very first GCs and calculated when the entire Galaxy became
ionised. When this occurred, we halted GC formation.

There is some uncertainty in our calculation of the amount of
ionisation from the first GCs. We have therefore treated this as a
free parameter which we have adjusted so that our model produces a
similar total number of metal-poor GCs to those observed in the Milky
Way. Even though this has been adjusted, we should emphasise that the
values chosen are quite reasonable and that changing the value by as
much as a factor of 2 will only vary the redshift when GC formation is
suppressed over the range of z$_{ion}$ = 10-15. Our adopted value of
the ionisation contribution (mass of baryons ionised per mass of
baryons in the halo forming each GC) is $F_{ion} = 210$. This results
in the galaxy being ionised by a redshift of z$_{ion}$ = 13 when 173
metal-poor GCs have formed.

We have tested this model for the effects of numerical resolution by
repeating the analysis for a lower-resolution simulation. We obtained
almost identical results between the two resolution runs, indicating
that the model is not biased by the simulation resolution.

Having fixed the number of GCs, we then find that the model
successfully predicts two independent properties of the metal-poor
GCs: their spatial distribution and formation ages of the GCs. The
radial distribution of the model GCs shown in
Figure~\ref{fig:MPspatial} is a very good match to that of the
observed Milky Way population. We obtain mean ages of 13.3 Gyrs,
consistent with observations by \cite{Salaris02}. The agreement of our
model with these two independent observations provides strong support
for our approach.

We can also estimate the final masses of the GCs if we introduce a
second variable parameter, the star formation efficiency in the
proto-clusters. We have adjusted this (again within reasonable values)
to match the observed masses of Galactic metal-poor GCs. Although the
mean mass has been set, we note that we predict a narrower {\em range}
of mass than observed.

\subsection{Metal Rich GCs}

Our approach to the formation of metal-rich GCs is much more
qualitative as the process is much more dependent on gas processes
that are not directly described in the current simulations. Our aim
was to find classes of model that could reproduce distributions like
those of Galactic metal-rich GCs which are much more
centrally-concentrated than the metal-poor GCs. This distribution
cannot be produced by hierarchical processes because the metal-rich
GCs are younger than the metal-poor GCs; in hierarchical processes it
is always the oldest objects that are most centrally-concentrated.

We tested two formation mechanisms: cluster formation triggered in the
gas disk of the central halo by large {\em merging} events and the
stripping of infalling haloes (presumed to contain GCs) as they merge
with the central halo. We rejected the stripping model because it
produces a spatial distribution of GCs in the present day that is far
too extended (see Fig. \ref{fig:mrprof}).

The only model that produced a sufficiently central distribution was
the merger model. In the model we consider that GCs form when galactic
haloes of mass larger than $8 \times 10^8$ M$_\odot$ that merge with a
larger halo. We then assume that the GC formation takes place in the
central gas disk at a radius estimated as 0.1 times the dark matter
half-mass radius. Although there are quite a few assumptions in this
model, this model produces the correct distribution.  The model also
predicts a large spread of ages (8 - 13 Gyrs) that is mostly
consistent with the observed Galactic age estimates of
\cite{Salaris02}.

As with the metal-poor GCs, we find that this model is not biased by
the simulation resolution: when we apply the model to the
lower-resolution simulation there is no observable change in the
radial distributions within 100\,kpc of the central halo.

\subsection{Future Work}
There are a number of avenues for future work. Similar analysis could
be carried out on the remaining five Aquarius haloes. They
have slightly different central halo masses and spatial resolutions,
so they will allow us to test the same formation mechanisms
across subtly different evolutionary environments. We have already
carried out similar calculations on the AqF Aquarius halo and found
comparable results.

The next major step is to introduce semi-analytic modeling of
\cite{Bower06}. This could give insight into how merger events and
halo accretion can alter GC properties. With respect to the metal-rich
GCs, the nature of the gas disk during each merger event could be
inferred from such models and used to more accurately determine where
in the Galaxy these star-forming regions will occur.

Future work will also include GC formation within the Millennium-II
(MII) simulation which, although it has a lower resolution
($m_p\sim6.9\times10^6$M$_\odot$), can still locate GC formation sites
with a minimum of 10 particles per GC. The scale of the MII
simulations (box-side length of 100 Mpc/h) will enable us to correlate
GC properties with those of their hosts in a wide range of physical
environments.

\section*{Acknowledgements}

We wish to thank Bill Harris for many valuable comments about the observational data.

The simulations for the Aquarius Project were carried out at the
Leibniz Computing Center, Garching, Germany, at the Computing Centre
of the Max-Planck-Society in Garching, at the Institute for
Computational Cosmology in Durham, and on the 'STELLA' supercomputer
of the LOFAR experiment at the University of Groningen.

This work was supported by travel funding from the Australian Research Council (grant LX0775963). PAT was partially supported by an STFC rolling grant. Brendan Griffen would like to acknowledge the support provided by the University of Queensland via a University of Queensland Postgraduate Scholarship.
\bibliography{biblio}
\section*{Appendix}
\label{app:qbar}

Here we estimate $\bar{q}$, the mean number of ioising photons
per baryon, averaging over a Salpeter initial mass function.  The
calculation is easily extended to other mass functions that can be
approximated as a power law above 10\,$M_\odot$.

The Salpeter IMF is
\begin{equation}
\mathrm{d}n={(\alpha-2)\,M\over m_0}\,\left(m\over
m_0\right)^{-\alpha}{\mathrm{d}m\over m_0},\hspace{1cm}m\geq m_0,
\end{equation}
where $\mathrm{d}n$ is the number of stars in the mass interval
$m\mapsto m+\mathrm{d}m$, $M$ is the total mass of stars, and
$\alpha=2.35$ and $m_0=0.1\,M_\odot$ are parameters describing the
slope and lower mass-cut of the population.

The number of ionising photons per baryon as a function of stellar
mass for a typical Population~2 metallicity of 0.001, derived using the
Starburst~99 code of \citet{Leitherer99}, is given in Figure~2 of
\citet{Tumlinson04}.  We approximate this a sequence of piecewise,
linear fits in $\lambda=\log_{10}(m/M_\odot)$:
\begin{equation}
q\approx\left\{\begin{array}{ll}
\ 30\,000\,(\lambda-1.00),& 1.00<\lambda<1.18\\
108\,000\,(\lambda-1.13),& 1.18<\lambda<1.72\\
\ 20\,000\,(\lambda+1.47),& 1.72<\lambda<2.10.
\end{array}\right.
\label{eq:q}
\end{equation}
\citet{Tumlinson04} do not give the value of $q$ for masses above
120\,$M_\odot$; there are few stars of this mass and so their
contribution to $\bar{q}$ is relatively small: we simply take
$q$ to be constant in this regime.

Within each piecewise interval the contribution to $\bar{q}$ is
\begin{eqnarray}
\bar{q}&=&\int_{m_1}^{m_2}q\,(\alpha-2)\left(m\over
m_0\right)^{1-\alpha}{\mathrm{d}m\over m_0} \\
&=&\int_{\mu_1}^{\mu_2}k_q\,(\alpha-2)\log_{10}(\mu/\mu_k)\,\mu^{1-\alpha}{\mathrm{d}\mu},
\end{eqnarray}
where $\mu=m/m_0$ and $k_q$ and $\mu_k$ are appropriate constants taken from
Equation~\ref{eq:q}.  This integrates to give
\begin{equation}
\bar{q}=k_q\,\left[\left(\log_{10}(\mu/\mu_k)+\log_{10}e/(\alpha-2)\right)\mu^{2-\alpha}\right]_{\mu_2}^{\mu_1}.
\end{equation}
This expression can be simplified by noting that, when summing these
expressions over the whole mass range, the first term in the square
brackets vanishes whenever $q$ is a continuous function.

Putting in values appropriate to a Salpeter IMF gives $\bar{q}\approx10\,000$.

\label{lastpage}

\end{document}